\documentclass[twocolumn,secnumarabic,amssymb,nobibnotes,aps, prl,superscriptaddress]{revtex4-1}

\usepackage{graphicx} % Include figure files
\usepackage{dcolumn}  % Align table columns on decimal point
\usepackage{bm}       % Bold math
\usepackage{hyperref} % Add hypertext capabilities
\usepackage{natbib}
\usepackage{amsmath}
\usepackage{amsfonts}

\begin{document}
	
\title{Self-Adaptive Amplified Spontaneous Emission Suppression with a Photorefractive Two-Beam Coupling Filter}

\author{Jacob Pettine}
\affiliation{JILA and Department of Physics, University of Colorado and National Institute of Standards and Technology, Boulder, CO 80309-0440, USA}

\author{Miao Zhu}
\affiliation{Agilent Laboratories, Agilent Technologies, 5310 Stevens Creek Blvd, Santa Clara, CA 95051}

\author{Dana Z. Anderson}
\affiliation{JILA and Department of Physics, University of Colorado and National Institute of Standards and Technology, Boulder, CO 80309-0440, USA}

%%%%%%%%%%%%%%%%%%%%%%%%% Abstract %%%%%%%%%%%%%%%%%%%%%%%%%%%%%%
	
\begin{abstract}

Amplified spontaneous emission is a source of broadband noise that parasitically limits the achievable gain in laser amplifiers. While optical bandpass filtering elements can suppress these broadband noise contributions, such filters are typically designed around specific frequencies or require manual tuning, rendering them less compatible with tunable laser systems. Here, we introduce a nonlinear self-adaptive filter and demonstrate the suppression of amplified spontaneous emission surrounding the lasing mode of a tunable 780 nm external cavity diode laser, using the two-beam coupling interaction in photorefractive BaTiO$_3$. A peak suppression of $-$10 dB is observed $\pm$2.5 nm from the lasing mode, with an overall 50\% filter power throughput. The dynamic photorefractive filter is automatically centered on the peak frequency due to the continuous writing and readout of the volume holographic grating and can thereby also automatically adapt to frequency tuning, drift, or mode hopping with an estimated auto-tuning rate of 100 GHz/s under typical conditions. Additionally, we present opportunities for enhancing filter suppression characteristics via the input intensity ratio and tuning the bandwidth via the coupling angle, toward versatile, self-adaptive optical filtering.

\end{abstract}
	
\pacs{1111}% PACS, the Physics and Astronomy
\maketitle

%%%%%%%%%%%%%%%%%%%%  Main Text %%%%%%%%%%%%%%%%%%%%%%%%%%%%%%

Photorefractive crystals have been utilized for a variety of spectral filtering applications, including carrier suppression for optical communications \cite{Anderson:2001}, cavity mode selection \cite{Maerten:2002}, laser bandwidth narrowing \cite{Chomsky:1992}, and narrow-band notch and/or bandpass filtering \cite{Rakuljic:1993,Herve:1994}. A key property of these materials is the efficient generation of charge density gratings and a large $\chi^{(2)}$ susceptibility to the corresponding DC or low-frequency space-charge fields, leading to efficient nonlinear interactions even for relatively weak continuous wave laser fields (i.e. intensities $<$ 1 W/cm$^2$). While previous photorefractive filtering schemes have primarily utilized either fixed gratings or dynamic gratings maintained by writing beams to filter a separate incident read beam, here we employ the two-beam coupling interaction in BaTiO$_3$ for dynamic self-filtering. In particular, we demonstrate broadband (few-THz) amplified spontaneous emission (ASE) power suppression surrounding the lasing mode of a tunable $\sim$780 nm external cavity diode laser, with the filter automatically centered on the lasing mode due to the continuous writing and readout of the grating.

The two-beam coupling interaction occurs between two temporally coherent beams propagating within a photorefractive crystal. As the two beams interfere within the crystal with a periodic intensity modulation, charge carriers excited in the bright regions diffuse to the dark regions, where they recombine with vacancy/impurity trap states within the insulator bandgap. The resulting periodic charge density and corresponding space-charge electric field then modulates the refractive index via the linear electro-optic (Pockels) effect, a $\chi^{(2)}(\omega=\omega+0)$ process in which the index modulation depends linearly on the static space-charge field \cite{Boyd:2008}. A volume phase grating is thus established via the photorefractive effect. For diffusion-limited carrier transport, the grating is spatially phase-shifted from the interference fringes by $\pm\pi/2$, with the sign depending on the crystal orientation \cite{Gunter:1982}. Due to this $\pi/2$ phase shift, the diffracted and transmitted beams interfere constructively along one output path and destructively along the other. Optical energy is thereby coherently transferred from one beam (the pump beam) to the other (the signal beam) via the two-beam coupling interaction, as illustrated in Fig. 1. 

\begin{figure}[b]
\includegraphics{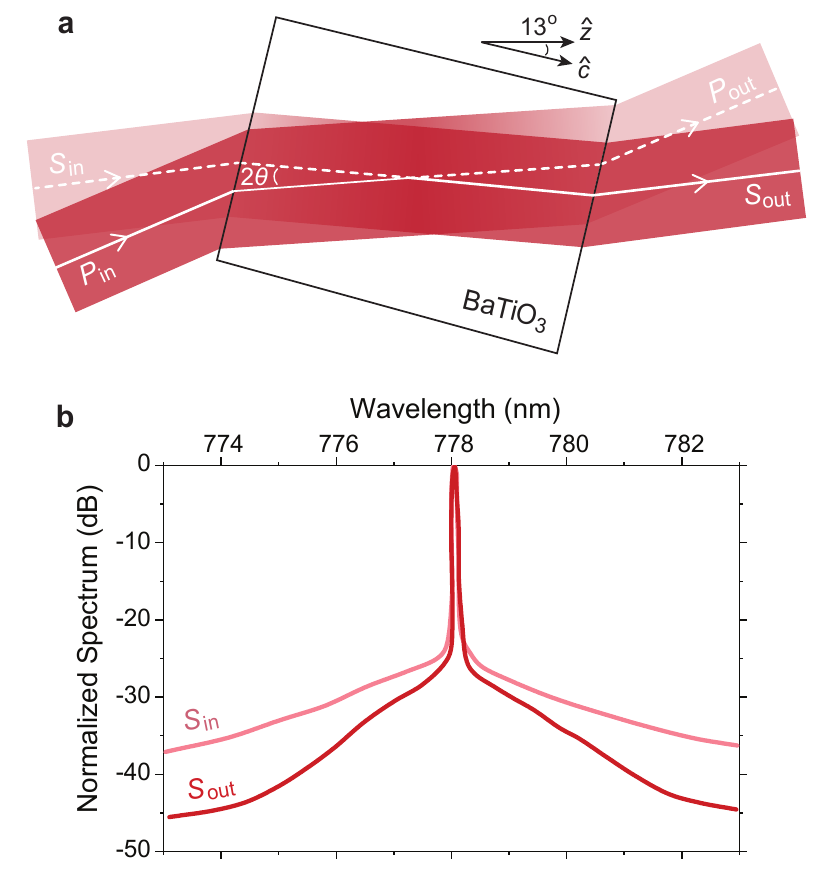}
\caption{(a) Illustration of the two-beam coupling configuration, with experimental angles and relative dimensions. (b) Example signal beam input and filtered output spectra, normalized to the peak and demonstrating ASE suppression.}
\label{Fig1}
\end{figure}

Modeling broadband ASE suppression requires a many-frequency two-beam coupling treatment \cite{Pecchia:1999}. The continuous pump and signal beam spectra are approximated as sums over plane waves in the coupled differential equations,
	\begin{align} \label{Eq_TwoBeamCoupling}
	\begin{split}
	\frac{dS_j}{dz} &= \frac{\gamma}{2I_0} P_j \sum_q S_q P_q^\ast 
	e^{i\left(\Delta k_{qj}z + \phi_{qj}\right)} -\frac{\alpha}{2}S_j ,\\
	\frac{dP_j}{dz} &=-\frac{\gamma}{2I_0} S_j \sum_q S_q^\ast P_q 
	e^{-i\left(\Delta k_{qj}z+\phi_{qj}\right)}  -\frac{\alpha}{2}P_j ,
	\end{split}
	\end{align}
in which $S_j$ and $P_j$ are the signal and pump electric field amplitudes for frequency component $j$, respectively, and $I_0 = \sum_q \left( |S_q(z=0)|^2 + |P_q(z=0)|^2 \right)$. The coupling constant, $\gamma$, is given by \cite{Yeh:1993}
	\begin{equation} \label{Eq_Gamma}
	\gamma = \frac{2\pi n_1}{\lambda_\mathrm{p} \cos\theta},
	\end{equation}
where $\theta$ is the internal half-angle between the beams (see Fig. 1), $\lambda_\mathrm{p} =$ 778 nm is the peak wavelength, and $n_1$ is the index modulation, which is proportional to $\cos 2\theta$ for the p-polarized beam interference. The absorption coefficient is measured to be $\alpha =$ 0.25 cm$^{-1}$, similar to previous measurements \cite{Klein:1986, Ducharme:1986}. The absorption is known to depend sensitively on the crystal preparation (trap density) \cite{Klein:1986} and laser intensity \cite{Brost:1988}, but merely serves to limit the maximum ASE suppression and total filter throughput without otherwise affecting the general filtering properties.

The exponential terms in Eq. \ref{Eq_TwoBeamCoupling} account for the Bragg-mismatched diffraction of frequency $\nu_j$ off of the grating established by frequency $\nu_q$, thereby entirely encoding the spectral filtering effect. The first term,
	\begin{equation} \label{Eq_kPhase}
	\Delta k_{qj} = \frac{4\pi n}{c}\frac{\sin^2\theta}{\cos
	\theta}\left(\nu_q-\nu_j\right),
	\end{equation}
depends upon the relative frequency detuning and $\theta$, where $n = 2.4$ is the average refractive index. The second phase-mismatch term \cite{Chomsky:1992},
	\begin{equation} \label{Eq_phiPhase}
	\phi_{qj} = \frac{2\pi}{c}\left(\nu_q-\nu_j\right)	
	\Delta l,
	\end{equation}
depends upon the relative frequency detuning and the optical path length difference between the signal and pump beams, $\Delta l$, which is positive for a longer signal arm. Eq. \ref{Eq_phiPhase} describes a spatial phase shift of the grating in the direction of the grating wavevector. Frequency $\nu_j$ therefore experiences a reduced diffraction efficiency off of the grating due to frequency $\nu_q$ if $\phi_{qj}$ is not an integer multiple of $2\pi$, as the grating phase shift is no longer at the optimal $\pi/2$ value and the interference between the diffracted pump and transmitted signal fields is no longer perfectly constructive. Unless otherwise specified, the condition $\Delta l = 0$ is maintained between the pump and signal beams to emphasize the primary filtering effect due to Eq. \ref{Eq_kPhase}.

Spectral filtering is due to the combination of (i) the nonlinear $\chi^{(2)}$ interaction, in which the grating strength is proportional to the intensity modulation, $\propto S_j P_j^\ast /I_0$, and (ii) the Bragg mismatch between different frequency components, where gain falls off with detuning from a particular grating frequency with a sinc functional dependence \cite{Yeh:1993}. Therefore, the strongest input frequency components around the laser peak experience the strongest two-beam coupling gain, while the off-peak gain is much weaker. This leads to less relative ASE in the signal beam output and thus to ASE suppression, as shown in Fig. 1. The spectral filtering is predominantly due to the $\Delta k$ phase mismatch term (Eq. \ref{Eq_kPhase}), while the $\phi$ term (Eq. \ref{Eq_phiPhase}) leads to additional spectral oscillations in the event of an optical path length mismatch. 

Filtering is achieved here with an unfixed grating, which is continuously written and read out such that the grating is automatically established for the appropriate laser frequency and also automatically adapts to changes in the laser spectrum, such as drift or mode-hopping. In particular, the filtering is largely unaffected by drift that remains within the Bragg-degeneracy region of the previously formed grating (approximately $\pm$10 GHz) within the time it takes for a new grating to form. The photorefractive response time of BaTiO$_3$ crystals is typically between 100 ms and 1 s for $I_0$ around 100 mW/cm$^2$ \cite{Ducharme:1984, Feinberg:1980}. An illuminated crystal with a response time of 100 ms would therefore track laser frequency drifts on the order of 100 GHz/s, or 0.2 nm/s around 780 nm, with minimal degradation to the coupling efficiency or the corresponding spectral filtering. Only the auto-peak-tuning is demonstrated explicitly here, but we note for further consideration that BaTiO$_3$ is known to have a rather slow response time, such that other photorefractive crystals may be considered for faster adaptability \cite{Gunter:2007}.

\begin{figure}[b]
\includegraphics{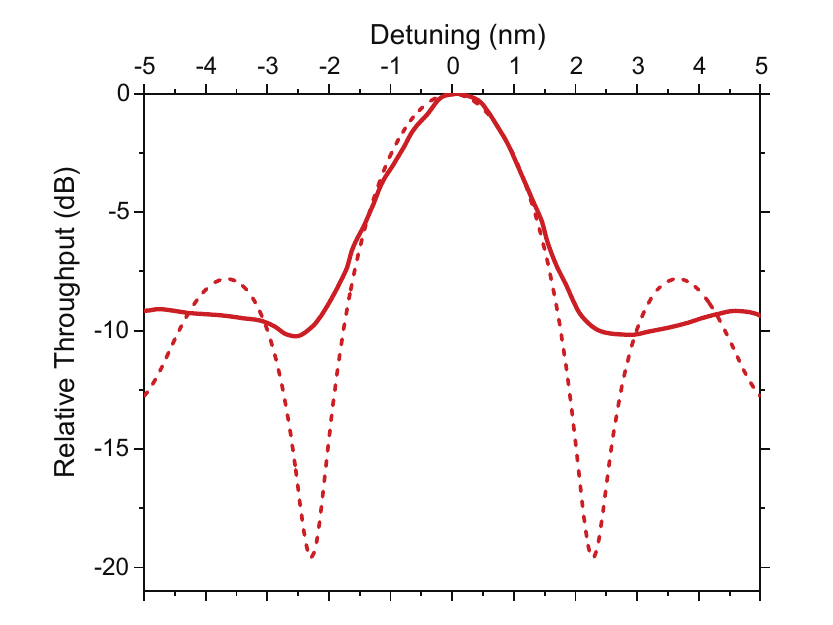}
\caption{Measured (solid) and calculated (dashed) filter throughput spectra relative to the laser peak at 778 nm for $r=$ 15 and $\theta=$ 4.5$^{\circ}$, with $-$10 dB ASE suppression observed at $\pm$ 2.5 nm detuning.}
\label{Figure2}
\end{figure}

The photorefractive BaTiO$_3$ used here is an undoped, uniaxial poled 5 $\times$ 5 $\times$ 7 mm ferroelectric crystal, with the $\hat{c}$ axis along the 7 mm dimension. Coupling is arranged in the transmission geometry (Fig. 1) with the beams extraordinary polarized to take advantage of the crystal's large $r_{42}$ electro-optic coefficient \cite{Gunter:2007}. To maximize the observable ASE suppression within the dynamic range of the optical spectrum analyzer, the diode operating current is reduced down to the lowest stable level, corresponding to a peak-to-ASE power ratio of 25 dB, with the ASE rolling off to either side of the peak by approximately $-$2.5 dB/nm before filtering. The total input power is set to 1 mW, for to a total intensity of $\sim$25 mW/cm$^2$ within the crystal given the 2 mm diameter beams, which are collimated (unfocused) for optimal spatial overlap. The interaction length within the crystal is approximately $L =$ 7 mm, and $\theta = 4.5^{\circ}$ unless otherwise stated. For these parameters, the coupling constant is determined by measuring the gain as a function of the input beam intensity ratio to be $\gamma =$ 5.5 cm$^{-1}$ (see Eq. \ref{Eq_PeakGain} below). From Eq. \ref{Eq_Gamma}, the value of the index modulation, $n_1$, is then $0.68 \times 10^{-4}$. We note that imperfect signal and pump mode spatial overlap within the crystal (illustrated in Fig. 1) is automatically compensated for in the ``effective" measured $\gamma$. 

\begin{figure}[t]
\includegraphics{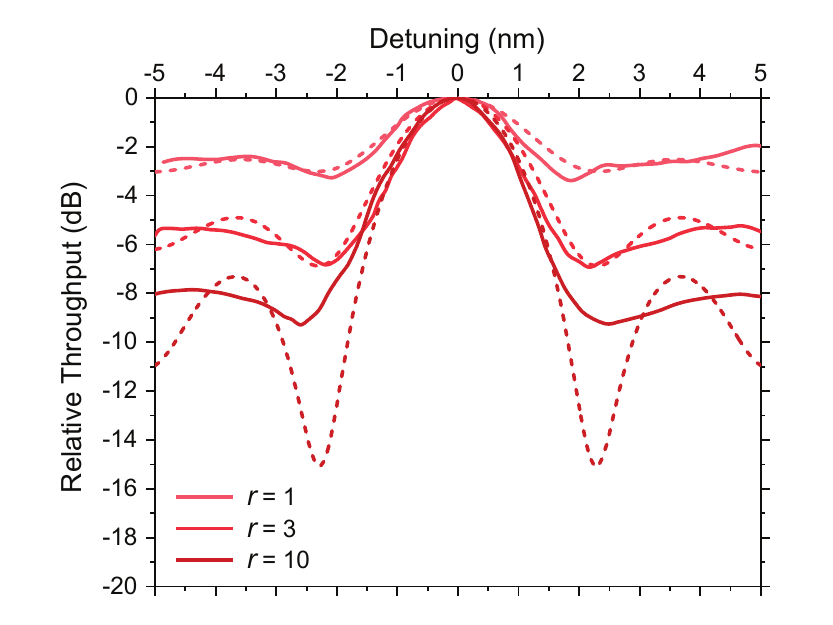}
\caption{Measured (solid) and calculated (dashed) filter throughput spectra at a series of input beam intensity ratios $r$.}
\label{Figure3}
\end{figure}

Experimental and theoretical spectral filtering curves are shown in Fig. 2, with the relative throughput defined as the ratio of the normalized signal beam output spectrum to the normalized input spectrum. A maximum suppression of $-$10 dB is observed at $\pm$ 2.5 nm detuning from the 778 nm peak, with a FWHM suppression bandwidth of 1.8 nm. It will be shown that the suppression and bandwidth depend sensitively on the input beam intensity ratio ($r$) and the coupling angle ($\theta$), respectively. The input intensity ratio,
	\begin{equation}
	r_j=\frac{\lvert P_{j}(z=0)\rvert^2}{\lvert S_{j}(z=0) 	
	\rvert^2},
	\end{equation}
is adjusted via half-waveplate and polarizing beamsplitter. Since the pump and signal beams originate from the same source and propagate with the same transfer functions prior to the crystal, $r_j=r$ for all frequency components. The measured overall two-beam coupling gain of 8 for $r=15$ yields a $\sim$50\% overall filter throughput efficiency, i.e. the ratio of the filtered signal beam output power to the total input power. 

A fourth-order Runge-Kutta method is implemented to integrate Eq. \ref{Eq_TwoBeamCoupling}, in which the experimental optical spectrum analyzer resolution (0.1 nm) is utilized as the sampling interval for the frequency components, and the measured input absolute power spectrum (as opposed to the power density spectrum) is thus taken as the initial spectrum in the calculations. The weaker suppression observed in the experimental curves relative to the theoretical predictions in Fig. 2 and elsewhere is primarily attributed to the limited dynamic range (optical rejection ratio) of the optical spectrum analyzer, which is approximately 35 dB within a $\pm$1 nm range and 45 dB within a $\pm$5 nm range. Thus, $-$10 dB is the observable limit for the input 25 dB peak-to-ASE power ratio. 

\begin{figure}[b]
\includegraphics{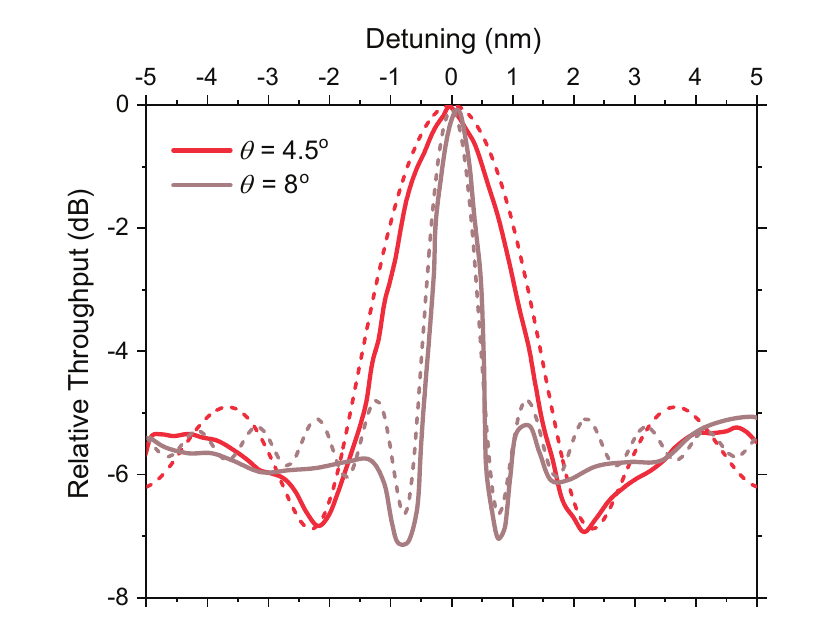}
\caption{Measured (solid) and calculation (dashed) filter throughput spectra for $r=$ 3 at two different internal beam coupling half-angles, $\theta$, demonstrating the predicted bandwidth narrowing for larger angles due to enhanced Bragg mismatch detuning sensitivity.}
\label{Figure4}
\end{figure}

Approximate theoretical limits on ASE suppression can be determined in a simple manner through the monochromatic coupling expression \cite{Yeh:1993}, with peak frequency gain,
	\begin{equation} \label{Eq_PeakGain}
	G_\mathrm{p}=\frac{\lvert S_p(z=L) \rvert^2}{\lvert S_p(z=0) 
	\rvert^2} = \frac{1+r}{1+re^{-\gamma L}} e^{-\alpha L}.
	\end{equation}
At large detuning from the peak, the ASE experiences essentially no gain, and therefore Eq. \ref{Eq_PeakGain} serves as an $r$-dependent limit on the ASE suppression. This limiting behavior already becomes apparent within the $\pm$5 nm detuning range in Fig. 3, in which the limiting suppression values are measured approximately at $-$2.5 dB, $-$5.5 dB, and $-$8 dB for $r=1$, $r=3$, and $r=10$, respectively. These are in reasonably good agreement with the $-$2.2 dB, $-$4.9 dB, and $-$8.4 dB values predicted by Eq. \ref{Eq_PeakGain}, with the full suppression curves calculated via Eq. \ref{Eq_TwoBeamCoupling} also yielding good agreement. By accounting for the $\theta$ dependence of $\gamma$ (Eq. \ref{Eq_Gamma}), Eq. \ref{Eq_PeakGain} also describes the effect of $\theta$ on the gain.

In addition to suppression tuning via $r$, the filter bandwidth can be tuned via the coupling angle, $\theta$, as demonstrated in Fig. 4. Bandwidth narrowing occurs for larger coupling angles due to the $\cos^2\theta/\sin\theta$ dependence of Eq. \ref{Eq_kPhase}, which enhances the frequency detuning sensitivity of the phase mismatch. However, even neglecting limitations imposed by the crystal geometry and Fresnel reflections, which may be overcome via custom crystal cutting \cite{Damiao:2001}, the beams become orthogonally polarized as $\theta \rightarrow \pi/4$ and the interference fringe contrast goes to zero, as encoded in the $\gamma \propto \cos 2\theta/\cos\theta$ dependence in Eq. \ref{Eq_Gamma}. Thus, there exists a trade-off between bandwidth and total suppression, which can be optimized for specific applications. 
  
\begin{figure}[t]
\includegraphics{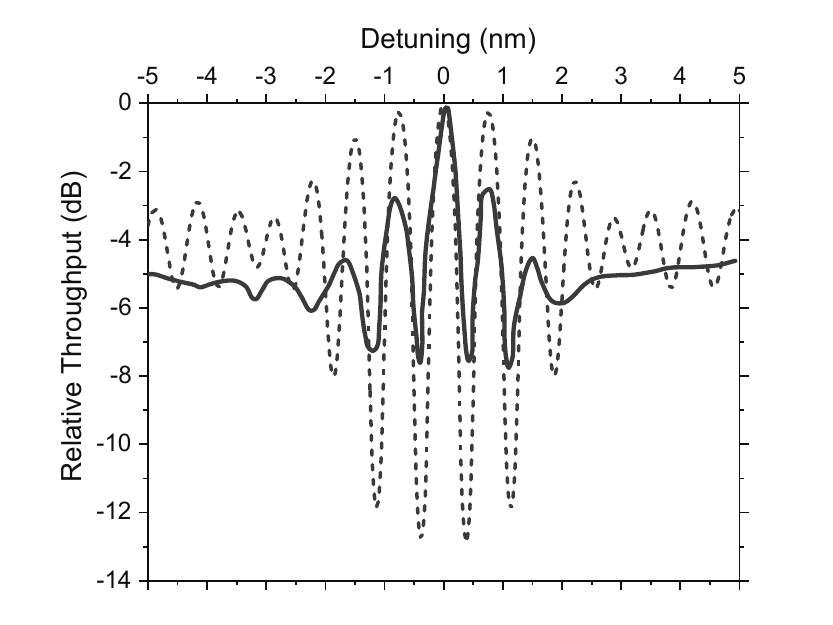}
\caption{Measured (solid) and calculated (dashed) filter throughput spectra for $r=$ 2 and $\theta=$ 4.5$^{\circ}$, with an optical path length of $\Delta l=0.9$ mm added to the signal arm relative to the pump arm, leading to oscillations due to the $\phi$ phase mismatch. }
\label{Figure5}
\end{figure}

Finally, the effects of optical path length differences between the signal and pump inputs in Eq. \ref{Eq_TwoBeamCoupling} are briefly examined, with further details described by Chomsky \textit{et al.} \cite{Chomsky:1992}. A $\Delta l=0.9$ mm optical path length delay is generated by placing a glass slide in the signal beam path. The resulting suppression curve is shown in Fig. 5, with the additional oscillatory behavior in good agreement with the theoretical predictions from Eq. 1. 
   
In summary, an auto-tuning optical bandpass filter has been demonstrated using the dynamic two-beam coupling interaction in photorefractive BaTiO$_3$ and utilized here to suppress ASE noise in a laser diode spectrum by at least $-$10 dB within $\pm$2.5 nm of the lasing mode. Larger suppression values (requiring more sensitive measuring equipment) are expected to occur for larger input intensity ratios, longer interaction lengths, and/or crystals with a larger coupling constant. For particular crystal cuts and coupling configurations, the filter bandwidth can be optimized by tuning the angle between the pump and signal beams. While ASE suppression serves as a suitable test of broadband spectral filtering, these effects could be tailored to a number of adaptive filtering applications for broad or multimode beams. Conversely, the opposite effect (carrier suppression) is achieved in the pump output \cite{Anderson:2001}. As a final note for practical implementation, this two-beam coupling filter can be miniaturized and/or fiber-coupled \cite{Damiao:2001}, and thus packaged as a passive optical element that can be incorporated into both free space and fiber-based laser systems, with manual or automated control knobs to adjust the filter spectrum via input intensity ratio and beam coupling angle.

\bigskip

%%%%%%%%%%%%%%%%%%%% Acknowledgements %%%%%%%%%%%%%%

This work was funded by Agilent University Relations and the NSF Physics Frontier Center (PHYS-1734006).

%%%%%%%%%%%%%%%%%%%% References %%%%%%%%%%%%%%%%%%%%%

\bibliography{PettineASESuppression_09_07_2020}

\end{document}